\documentclass[amsmath,amssymb,floatfix,prl,epsf,superscriptaddress,twocolumn]{revtex4}
\usepackage{amsmath,amssymb,natbib,bm,graphicx,url,epsfig,wasysym,mathrsfs,dsfont}
\usepackage[ansinew]{inputenc}
\usepackage{bm}
\usepackage{color}
\usepackage{ bbold }
\usepackage[fleqn,tbtags]{mathtools}
\newcommand{\be}{\begin{equation}}
\newcommand{\ee}{\end{equation}}
\newcommand{\bes}{\begin{equation}\begin{split}}
\newcommand{\ees}{\end{split}\end{equation}}
\newcommand{\bea}{\begin{eqnarray}}
\newcommand{\eea}{\end{eqnarray}}
\newcommand{\nn}{\nonumber}

\def\beq{\begin{equation}}
\def\eeq{\end{equation}}
\def\bea{\begin{eqnarray}}
\def\eea{\end{eqnarray}}
\def\ena{\end{eqnarray}}

\begin{document}


\title{Topological spin ordering via Chern-Simons superconductivity}
\author{Tigran Sedrakyan}
\affiliation{William I. Fine Theoretical Physics Institute,  University of Minnesota, Minneapolis, Minnesota 55455, USA}
\affiliation{Physics Frontier Center at the Joint Quantum Institute, University of Maryland, College Park, Maryland 20742, USA}

\author{Victor Galitski}

\affiliation{ Joint Quantum Institute and Condensed Matter Theory Center, Department of Physics,
University of Maryland, College Park, Maryland 20742-4111, USA}
\affiliation{School of Physics and Astronomy, Monash University, Melbourne, Victoria 3800, Australia}

\author{Alex Kamenev}
\affiliation{William I. Fine Theoretical Physics Institute,  University of Minnesota, Minneapolis, Minnesota 55455, USA}

\begin{abstract} 
We use the Chern-Simons (CS) fermion representation of $s =1/2$ spin
operators to construct  topological,  long-range magnetically ordered states of
interacting two-dimensional  (2D) quantum spin models. We show that
the fermion-fermion interactions mediated by the dynamic CS flux
attachment may give rise to Cooper pairing of the fermions.
Specifically, in an XY model on the honeycomb lattice, this
construction leads to a ``CS superconductor," which belongs to a
topologically non-trivial in 2D symmetry class DIII, with
particle-hole and time-reversal symmetries. It is shown that in the
original spin language, this state corresponds to a symmetry protected
topological state, which coexists with a magnetic long-range order. We
discuss physical manifestations of the topological character of the
corresponding state and requirements for models that could host it.
\end{abstract}

\date{\today}

\maketitle

 {\it Introduction --} 
 Two-dimensional quantum spin models is a fascinating subject, which continue to attract attention of theoreticians and experimentalists alike\cite{arov,chakr,manous,RS,AA,mis,hon,rmp,SS,starykh,collab}.  What makes it particularly challenging from the theory standpoint  is the absence of a simple weakly interacting picture and controlled theoretical tools to describe the plethora of  possible ground states where strong quantum fluctuations abound.  Much of the earlier theoretical work in quantum magnetism has focused on long-range-ordered magnetic  phases, usually well-described in terms of the Schwinger boson representation of the spin operator, with subsequent employment of a mean-field theory or other methods (e.g., large-$N$ approaches and variational analyses)~\cite{arov,RS,SS,AA}. 

Another prominent class of ground states are  spin liquids, which have received much attention since the early nineties, boosted by the discovery of high-temperature superconductivity and some of its exotic scenarios\cite{ander87,kalmeyer,zee,wen-sl}. 
A hallmark of most spin liquids is a lack of a long-range order and a local order parameter. The theoretical description of these states often involves fractionalization - where the spin operators (or equivalently the operators of hardcore bosons) are represented as a product of two fermions (``partons''), which can ``fall'' into various mean-field states. This construction often leads to gauge theories, non-locality, and topological order of the underlying quantum liquid \cite{toric,fisher,palee,balents,kagome,potter,grov,sheng,essin,chiral,hermele,sava,top-class}.  By now, these kinds of spin liquids have been thoroughly classified, and there is a promising experimental evidence for their   actual existence in solid-state materials \cite{herbert,norman}.

Very recently, there has been a tremendous progress in identifying and classifying symmetry-protected topological (SPT) phases of interacting fermionic\cite{Altland,schnyder,K-theory,Zirnbauer,Chiu,gu,top-class} and bosonic\cite{wen1,levin,wen-sci,senthil,top-class} systems. The SPT phases have some properties of short-range entangled trivial phases, but are also distinct from those, e.g., by exhibiting edge modes. Hence, they in effect represent  a third class of possible ground states of strongly-correlated systems, including quantum magnets. In this paper, we propose a  microscopic technical construction that appears to give rise to  exotic states  of this latter type (and their ``gauged'' versions)  in interacting  lattice  spin models. 

A particularly simple example of an SPT spin phase was proposed by  Levin and Gu \cite{levin}, who considered the Ising paramagnet on a triangular lattice with the deceptively simple Hamiltonian $\hat{H}_{LG} = - \sum\limits_{{\bf r} \in \triangle}  \hat{S}_{\bf r}^x$, where the spin operators are either Pauli matrices $\hat{S}_{\bf r}^x = \frac{1}{2}\hat{\sigma}_{\bf r}^x$ (which indeed makes the corresponding phase a trivial Ising paramagnet) or $\hat{S}_{\bf r}^x \equiv \hat{B}_{\bf r}^x = \frac{1}{2}\hat{\sigma}_{\bf r}^x \prod_{( {\bf r}' {\bf r}'' )} \exp\left[ \frac{i}{4} \left(1 - \hat\sigma_{{\bf r}'}^z \hat\sigma_{{\bf r}''}^z\right)\right]$, where we use Levin-Gu notations for "twisted" spin operators with ${\bf r}'$ and ${\bf r}''$ running over the six triangles containing site ${\bf r}$. These operators satisfy the usual $su(2)$ algebra's commutation relations  and give rise to a distinct SPT phase, with non-trivial edge physics. To motivate the central question of this paper, we note that the Levin-Gu topological Ising model may appear in an interacting quantum spin model where the symmetry is broken either ``externally'' or  spontaneously, e.g., a twisted $XY$-model on a triangular lattice, $\hat{H}_{TXY} = - \sum\limits_{\langle {\bf r}{\bf r}' \rangle \in \triangle} \hat{B}_{\bf r}^+ \hat{B}_{{\bf r}'}^-$.  Indeed, the mean-field ordered state (e.g., with  the mean-field magnetization along the $x$-direction) essentially reproduces the SPT Ising model above $\hat{H}_{TXY} = - \langle \hat{B}^x \rangle \sum\limits_{ {\bf r} \in \triangle}  \hat{B}_{\bf r}^x$. Note that edge excitations and fluctuation effects in the ``$TXY$-model''  may lead to qualitative changes in the nature of the mean-field topological (SPT-like) phase, but the above simple construction does suggest that there exist {\em topologically non-trivial long-range-ordered states} of interacting quantum magnets. 

 {\it Chern-Simons fermionization --} 
 This paper provides an example and effective description of a topological long-ranged-ordered state of a quantum spin model. We will focus on a specific Hamiltonian  -- see, Eq.~(\ref{H})  below -- but the general method we use works for a wide class of lattice models and is based on the Chern-Simons (CS) flux attachment~\cite{AMP,Jain,EF,composite,Witten,Wen,we,chiral,motrun} - the Jordan-Wigner-type transform that ``converts'' hardcore bosons/spins into fermions  via attaching a string to each particle:
\begin{equation}
\label{CSt}
\hat{S}^{\pm}_{\bf r}=\hat{f}^{\pm}_{\bf r} \hat{\mathcal{U}}_{\bf r}^{\pm},\quad \hat{\mathcal{U}}_{\bf r}^{+}=\exp \left[ i e \sum_{{\bf r'}\neq{\bf r}}\arg({\bf r}-{\bf r'})\hat{n}_{\bf r'}\right].
\end{equation}
Here $\hat{S}^{\pm}_{\bf r}$ are the spin-$1/2$ raising/lowering operators on a lattice cite ${\bf r}$, $\hat{n}_{\bf r} = \hat{S}^z_{\bf r} + 1/2 = \hat{f}^+_{\bf r}\hat{f}_{\bf r}$, the sum runs over all lattice sites except ${\bf r}$, and $e$ is an odd integer CS charge, which makes $\hat{f}^{\pm}_{\bf r}$ into the fermion creation/annihilation operators. The resulting theory depends on a Hamiltonian and a lattice of course, and generally takes the form that is not amenable to an exact treatment. However, the theory - fermions coupled to the CS gauge field resulting from transformation (\ref{CSt}), 
\begin{equation}
\label{A} 
i \hat{\mathcal{U}}_{\bf r}^{+}\partial_{\mu}\hat{\mathcal{U}}_{\bf r}^{-}\rightarrow { A}_{\mu}({\bf r})= \varepsilon_{\mu\nu}\sum_{{\bf r'}\neq {\bf r}} \frac{({\bf r}-{\bf r'})_\nu}{|{{\bf r}}-{{\bf r'}}|^2}\, n_{\bf r'}
\end{equation}
 (with $\mu,\nu=1,2$, and $\varepsilon_{\mu\nu}$ being an antisymmetric tensor) provides a convenient field-theoretic platform to formulate an effective description  of various stable  phases of quantum magnets. 



These constructions usually proceed as follows. The CS gauge potential   is represented in terms of a  mean-field part (assumed static in the Lagrangian formulation) and fluctuations around the mean-field, ${\bf A} = \langle {\bf A} \rangle +\delta {\bf A}$. The fermions are integrated out on the background of the mean-field configuration (to be determined a posteriori via a variational analysis). Notice that in this construction  the CS fermions are assumed to simply fill up the single-particle bands (albeit with a non-trivial Hofstadter-type energy landscape) without undergroing a phase transition. The remaining low-energy theory - an expansion in the CS fluctuations, $\delta {\bf A}$ - provides a  field-theoretical description of the underlying mean-field. This way one can obtain various  states - both exotic and ordered ones. For example, an integer quantum Hall state of fermions generates a CS term, which can either add up to the statistical Chern-Simons field originating from transform (\ref{CSt}) (this corresponds to a chiral spin liquid) or cancel it  with the remaining Maxwell term representing a gapless phonon~\cite{EF} (this corresponds to an ordered state).

This elegant approach is not without its downsides. Just about any mean-field Ansatz for $\langle {\bf A} \rangle$  ``accidentally'' breaks physical symmetries that one may want to preserve. Furthermore, the CS fermions are actually not free, but rather represent strongly interacting entities. These interactions may lead to instabilities and hence new underlying spin phases. In this paper, we propose such an alternative construction of a topological long-range-ordered spin state via the CS flux attachment, where instead of assuming a specific mean-field for the CS gauge field, we treat it non-perturbatively as an interaction between the fermions that are shown to become unstable against pairing. 

 {\it The model --} 
 The specific model we use as our starting point is the bulk spin-1/2 antiferromagnetic Hamiltonian on the honeycomb lattice with nearest-neighbor couplings:
\begin{equation}
\label{H}
\hat{H}= J \sum_{\langle {\bf r} {\bf r}' \rangle \in \hexagon}\left[(1+\gamma) \hat{S}^{x}_{\bf r}\hat{S}^{x}_{{\bf r}'} +(1-\gamma)\hat{S}^{y}_{\bf r}\hat{S}^{y}_{{\bf r}'}\right],
\end{equation}
We emphasize that the purpose of our theory below is not to ``solve'' the particular model (in the sense of finding its lowest energy ground state, whose properties in the conventional setting are well known),  but to  illustrate that the appearance of topologically non-trivial long-range-ordered states is possible in a class of models. The ease and naturalness with which the  calculation goes through strongly suggests that this approach is generic in bipartite lattices (a similar calculation for a different model on the square lattice will be presented in a subsequent publication). 
Eq.~(\ref{H}) describes a 2D anisotropic XY-type model, whose bulk supports an antiferromagnetic ($J>0$) ground state. At $\gamma>0$ it corresponds to a doubly degenerate gapped phase  with N\'{e}el oder parameter  $\langle \hat{S}^{x}_{\bf r}\rangle$  with  $\mathds{Z}_2$ Ising symmetry which in this case is equivalent to reflection. 


In the absence of a net magnetization, CS fermionization yields a half filled fermionic system.  The Fermi level of fermions on the honeycomb lattice consists of two Dirac points conventionally denoted by $K$ and  $K'$.  Using the fermion representation of Eq.~(\ref{H}), and upon expansion in the vicinity of these Dirac points (below, we present calculation details for $\gamma=0$;  for a finite $\gamma$  the calculation is  essentially similar)  a gauge transformation generates the covariant derivative $\partial_\mu-ie{ A}_\mu({\bf r})$ (and kinetic momentum). The CS gauge field ${ A}_\mu$, that enters into the kinetic term, is bilinear in fermion operators and thus generates a two-particle interaction vertex. This brings the following momentum space representation of the Hamiltonian (\ref{H}):
$\hat{H} = \hat{H}_0 + \hat{H}_{int}$, where
\bea
\label{Dirac}
\hat{H}_0&=& v_F \sum_{\bf k}\big[\hat{f}^+_{\bf k,\alpha} {\bf k}\cdot {\bm \sigma}_{\alpha\beta} \hat{f}_{\bf k,\beta}-
\hat{\bar{f}}^+_{\bf k,\alpha} {\bf k}\cdot {\bm \sigma}^T_{\alpha\beta} \hat{\bar{f}}_{\bf k,\beta}\big]\\
\hat{H}_{int}&=&- \sum_{\bf k,k',q} V^{\alpha\alpha',\beta\beta'}_{\bf q} \hat{f}^{+}_{{\bf k},\alpha}\hat{\bar{f}}^{+}_{{\bf k'+q},\alpha'}
\hat{\bar{f}}_{{\bf k'},\beta} \hat{f}_{{\bf k+q},\beta'} .\nn
\ena
Here $v_F=\frac{\sqrt{3} J \varepsilon}{2}$ is the velocity at the Fermi level, $\varepsilon$ is lattice constant of the two triangular sub-lattices,
$\hat{f}^{\pm}_{ \bf k, \alpha}$ and  $\hat{\bar{f}}^{\pm}_{\bf k, \alpha}$ are low energy fermions with momenta measured from  $K$ and $K'$ points respectively,
and spinor indices correspond to the sub-lattices $\alpha= A,B$.  The interaction vertex $V$ in Eq.~(\ref{Dirac}) reads 
\bea
\label{HS}
V^{\alpha\alpha',\beta\beta'}_{\bf q}=2 \pi i e v_F \epsilon_{\mu\nu}\left(\sigma^\mu_{\alpha\beta}\delta_{\alpha'\beta'}+
\delta_{\alpha\beta}[\sigma^{\mu}]^T_{\alpha'\beta'}\right) A^\nu_{\bf q},
\ena
where ${\bf A_q}={\bf q}/|{\bf q}|^2 $ is the Fourier image of the  vector potential of the vortex gauge field, $\delta_{\alpha\beta}$ is the Kronecker delta symbol, and the summation over repeating indices is implied.

It is worth noting here that our fermionic Hamiltonian~(\ref{Dirac}) in momentum representation 
consists of graphene-like kinetic energy term $\hat{H}_0$ and non-local two-particle 
interactions that arise from integrating out the vortex operators in the fermionized representation. 
Aside from expanding in the vicinity of $K$ and $K^\prime$ points of the Brillouin zone, the above procedure is formally exact.


 {\it Cooper pairing of Chern-Simons fermions --} 
 To proceed further we make use of the Hubbard-Stratonovich transformation based on Cooper pair operators $\hat{\bar{f}}_{-\bf k,\alpha} \hat{f}_{ \bf k,\alpha'}$ and
 $\hat{f}^+_{\bf k,\alpha}\hat{\bar{f}}^+_{ -\bf k,\alpha'} $ to decouple four fermion interaction term in the Hamiltonian (\ref{Dirac}).
By introducing fluctuating superconducting order parameter fields $\Delta^{\alpha\beta}_{\bf k}$ we obtain 
\bea
\label{Hamiltonian}
H&=&H_0 +\sum_{\bf k}\hat{f}^+_{ \bf k,\alpha}\hat{\bar{f}}^+_{-\bf k,\alpha'}\Delta^{*\alpha\alpha'}_{\bf k}
+\Delta^{\alpha\alpha'}_{\bf k}\hat{\bar{f}}_{-\bf k,\alpha }\hat{f}_{ \bf k,\alpha}\big]\nn\\
&+&\sum_{\bf k,k'}\Delta^{*\alpha\alpha'}_{\bf k} \big[V^{-1}\big]_{\bf k-k'}^{\alpha\alpha',\beta\beta'}\Delta^{\beta\beta'}_{\bf k'},
\ena
where $V^{-1}$ is the inverse of the interaction vertex (\ref{HS}). Integrating out the fermionic degrees of freedom in Eq.~(\ref{Hamiltonian}) define an effective action $W(\Delta^{\alpha\beta}_{\bf k})$ for 
the superconducting order parameter $\Delta^{\alpha\beta}_{\bf k}$. We treat the latter in the stationary field approximation, similarly to the standard BCS theory of Cooper pairing. The corresponding saddle point equations,    
$\delta W/\delta\Delta^{\alpha\beta}_{\bf k}=0$, lead to: 
\bea
\label{gap}
\Delta^{\alpha\alpha'}_{\bf k}=
\sum_{\beta\beta' \bf k'}V^{\alpha\alpha',\beta\beta'}_{\bf k-k'}\langle \hat{\bar{f}}_{ -\bf k', \beta} \hat{f}_{ \bf k',\beta'}
\rangle.
\ena
Since the vertex function $V^{\alpha\alpha',\beta\beta'}_{\bf k-k'}$ in this expression is sharply momentum dependent,  the order parameter 
$\Delta^{\alpha\alpha'}_{\bf k}$ also turned out to  
be momentum dependent:
$\Delta^{11}_{\bf k}=\Delta^{22}_{\bf k}=\Delta_{3 {\bf k}}$,
$\Delta^{12}_{\bf k}=-\Delta^{21}_{\bf k}=\Delta^x_{0 {\bf k}}-i \Delta^y_{0 {\bf k}} $.
Here we have a vector order parameter ${\bm \Delta}_{0 {\bf k}}=(\Delta^x_{0 \bf k},\Delta^y_{0 \bf k}) $ and a rotation scalar $\Delta_{3 \bf k} $.
The latter corresponds to the pairing of fermions residing on the same sublattice, in contrast to the conventional BCS pairing, where particles 
having the same spin (here, instead of spin we have a pseudospin degree of freedom associated with two sublattices) do not get paired.
The lowest energy solution corresponds to $p$-wave pairing\cite{Volovik,Read,bern}; namely we look for
the momentum dependence of $\boldsymbol{\Delta}_{0 {\bf k}}$ in the form
$\boldsymbol{\Delta}_{0 {\bf k}}=\Delta_{0 {\bf k}} {\bf k}/k$, where $\Delta_{0 {\bf k}} = |\boldsymbol{\Delta}_{0 {\bf k}}|$ .
Substituting the solution (\ref{gap}) into Eq.~(\ref{Hamiltonian}) we obtain the Bogoliubov-de Gennes (BdG) Hamiltonian.
In the basis of 4-spinors $\psi_{\bf k}=(\hat{f}^A_{\bf k}, \hat{f}^B_{\bf k}, \hat{\bar{f}}^{A+}_{-\bf k}, \hat{\bar{f}}^{B+}_{-\bf k})$,
it acquires the form
\bea
\label{Hamiltonian-2}
H_{BdG}=\left(
\begin{array}{cc}
v_F{\bf k}\boldsymbol{\sigma} &\hat{\Delta}_{\bf k}\\
\hat{\Delta}^\dagger_{\bf k} & -v_F{\bf k}\boldsymbol{\sigma} 
  \end{array}
\right),
\ena
where  $\hat{\Delta}_{\bf k}= \Delta_{3 \bf k} \mathbb{1} + \boldsymbol{\Delta}_{0 \bf k} \times \boldsymbol{\sigma}$ and $\mathbb{1}$ is the identity matrix. $H_{BdG}$
gives a 4-band gapped spectrum  $\pm E^{(a)}_{\bf k}$, with the quasiparticle energy 
$E^{(a)}_{\bf k}=\sqrt{|a v_F {\bf k}+ \boldsymbol{\Delta}_{0 \bf k}|^2+|\Delta_{3 \bf k}|^2}$ 
and $a=\pm$ distinguishing between two upper/lower bands. As we see the spectrum is $U(1)$ rotationally invariant. 

\begin{figure}[t]
\includegraphics[width=40mm,angle=0,clip]{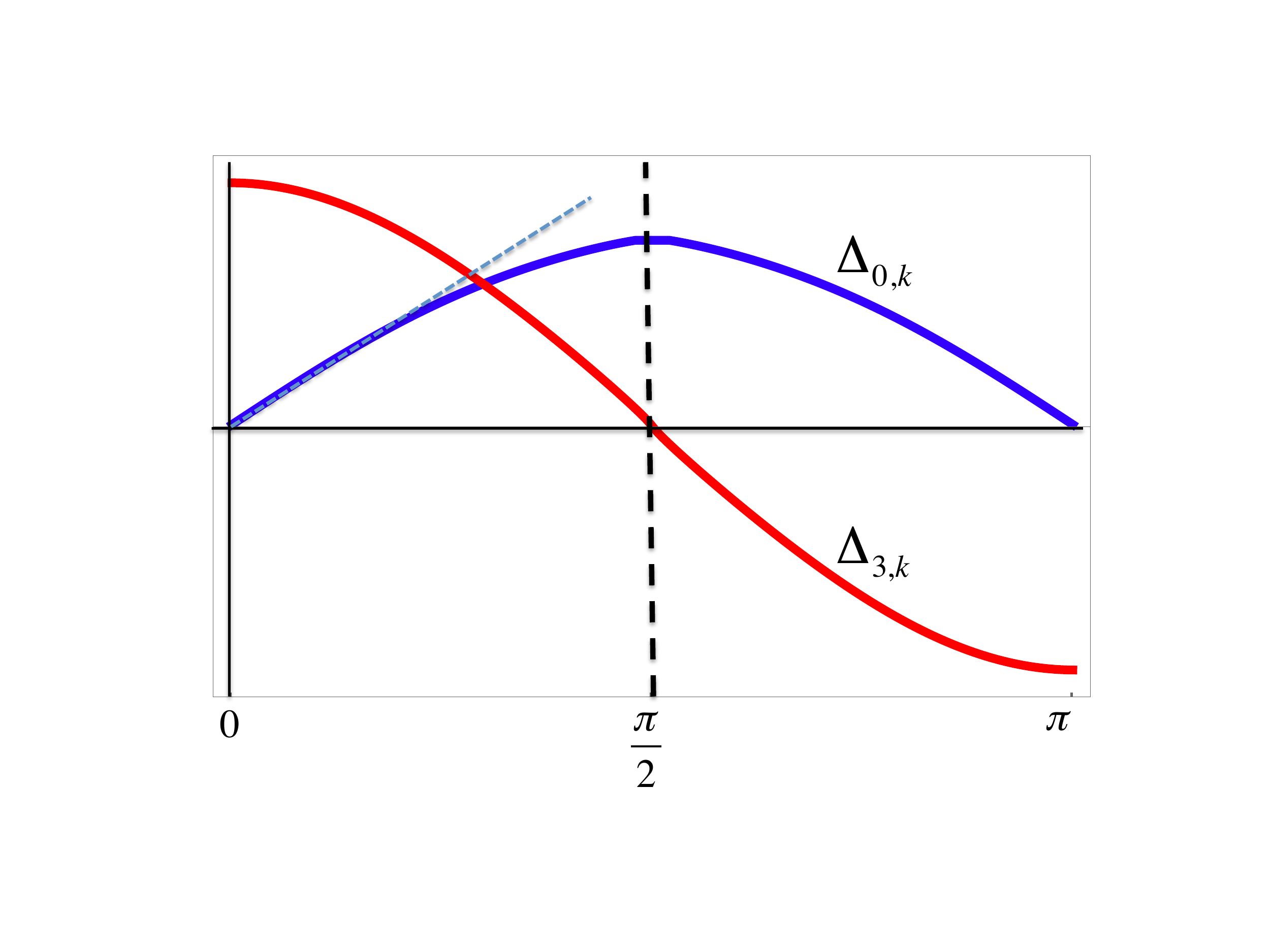}
\includegraphics[width=44mm,angle=0,clip]{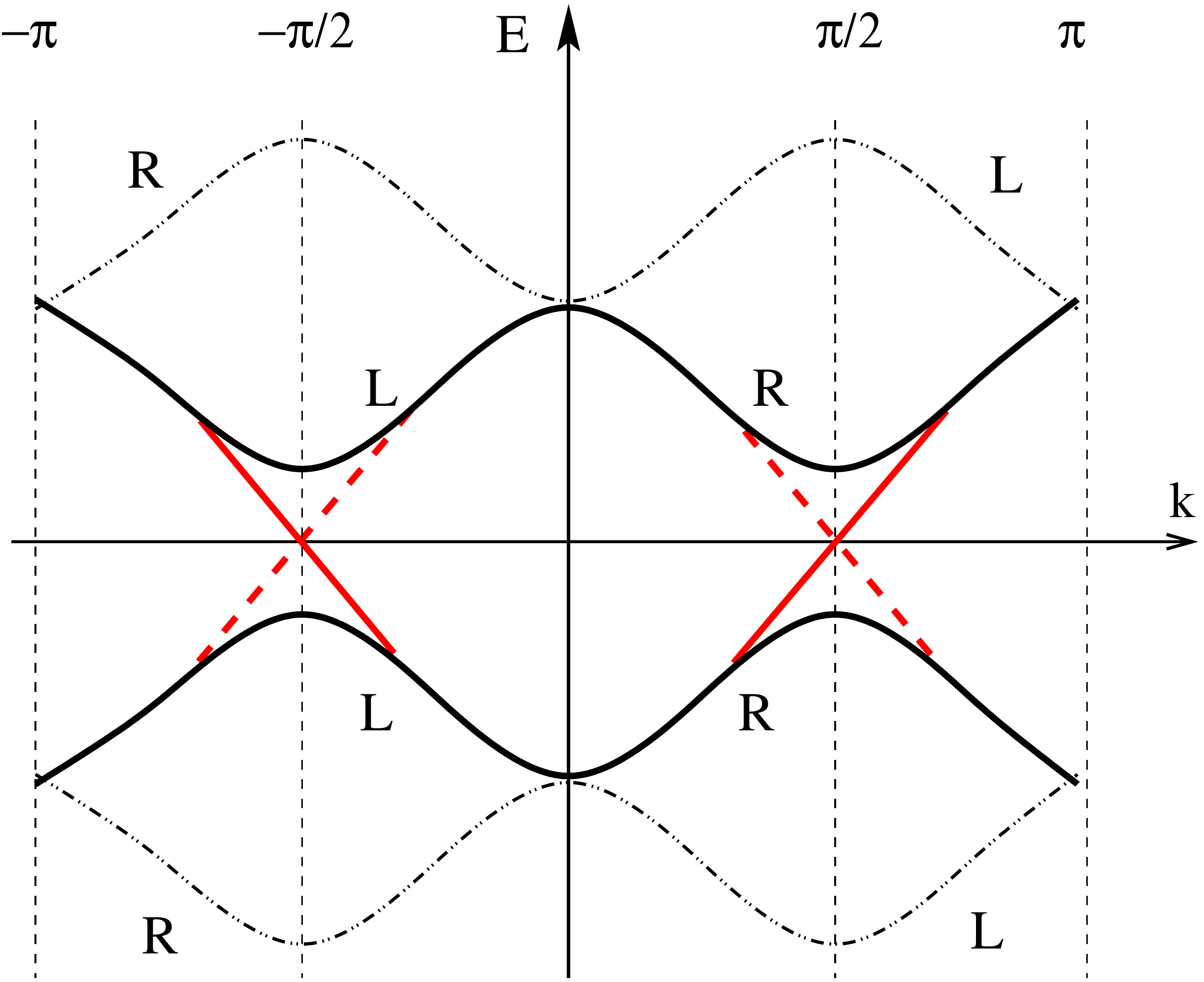}
\caption{(Color online) (a) Numerical solution of  gap equations (\ref{gap-eq}) for e=3 shown in arbitrary units. 
Cutoff parameter is $\Lambda\sim\pi/(2\varepsilon)$. (b) Thick black lines represent the low lying part of the bulk spectrum, dotted black lines represent the higher branches. R and L mark the branches corresponding to R/L states, see Eq.~(\ref{Hamiltonian-3}). Full red lines show R and L brunches of edge state energies. 
} 
\label{gap-graphic}
\end{figure}

To proceed, one needs to solve the self-consistency Eqs.~ (\ref{gap}) for the order parameters. 
Replacing the sum over 2D momenta by an integral and performing the angular integration 
one arrives at (for the details see the supplementary material) 
\bea
\label{gap-eq}
\Delta_{0 \bf k}&=&\frac{e v_F}{2}\sum_{a=\pm }\int_0^{k} dk' \frac{k' \Delta_{3 k'}}{k E^{(a)}_{\bf k'}},\nn\\
\Delta_{3 \bf k}&=&\frac{e v_F}{2}\sum_{a=\pm }\int_{k}^{\Lambda} dk' \frac{ \Delta_{0 k'}+a v_F k'}{ E^{(a)}_{\bf k'}},
\ena 
where we have  introduced a cutoff parameter $\Lambda$ around $K$(or $K'$) points defined by the area 
of the half of Brillouin zone (BZ). Solutions of Eqs.~(\ref{gap-eq})  in both halfs of BZ should be
glued with each other on the boundary to recover periodicity of the spectrum (see Fig.\ref{gap-graphic}).

The solution of Eq.~(\ref{gap-eq}) depends on  the CS charge, which in the case of  Hamiltonian (\ref{H})  must be an odd integer $e=1,3,5\cdots$. Remarkably, the simplest choice of $e=1$ yields only a trivial solution, with zero order parameter. However, the states with $e \geq 3$ (mathematically a solution exists for any  $e > e_c$, where the critical $e_c=2/3^{1/2}$; see the supplementary material) give rise to a nontrivial, gapped solution to Eq.~(\ref{gap-eq}), indicating that the CS fermions  are unstable
against pairing. The CS charge, $e$, is determined by energetics of a particular model and we found that  $e= 3$ yields the lowest energy (interestingly,  an analysis of nearest-neighbor spin-spin correlators  in our exotic state is quite close to those in the actual ground state of the  conventional $XY$ model~\cite{Campbell}).
 
Numerically found gap-functions for the CS superconductor are plotted in Fig.~\ref{gap-graphic}. We see, that $\Delta_{0,\bf k}$
is linear at $k<J/v_F$: $\Delta_{0,\bf k}\simeq\frac{3 v_F k}{2} $, as follows from the first of Eqs.~(\ref{gap-eq}).
This asymptote corresponds to the solution of the gap equation in one half of BZ, (e.g., around the $K$ point, i.e. on the segment $(0, \Lambda)$ of the momentum axis).
By flipping  signs of $\Delta_{0,\bf k}$ and $\Delta_{3,\bf k}$ in gap equations (\ref{gap-eq}), we generate a solution with opposite chirality. 
This is the solution in the vicinity of $K^\prime$.  

 {\it Magnetic long-range order --} 
 Now we turn to the discussion of the properties of the Chern-Simons superconducting state. First, we prove that the corresponding magnetic state has off-diagonal long-range order
and an associated local order parameter. The most natural correlation function to look at is the spin-spin 2-point correlation function, which in the fermion language takes
the form: $\left\langle \hat{S}_{\bf r}^+ \hat{S}_{\bf 0}^- \right\rangle = \left\langle \hat{f}^\dagger_{\bf r} \hat{f}_{\bf 0} e^{i \Phi_{\bf r}} \right\rangle$, where $\Phi_{\bf r}
= e \sum\limits_{{\bf r}'}  \left[ {\rm arg}\, {\bf r}' - {\rm arg}\, \left( {\bf r} - {\bf r}' \right)\right]$. The existence of the non-local string makes the calculation of the correlator complicated (we have not been able to evaluate it). However, one can construct high-order correlators, where the string effectively disappears and the spin and fermion correlators are one-to-one related. Note that the CS transform is one of infinitely many Jordan-Wigner-type fermionization transforms, that attach  strings in different ways through the lattice.  The observables (such as spin-spin correlators) must not depend on the ``gauge'' choice of a specific Jordan-Wigner string. One can check that there exists a choice of the string such that the four-point spin correlator - see Eq. (\ref{4point}) - is identically equal to the corresponding fermion one. On the other hand, the fermion correlation functions corresponding to two different Jordan-Wigner choices differ only by a phase and hence we arrive at the following relation
\begin{equation}
\label{4point}
C^{(4)}({\bf r} - {\bf r}') = \langle \hat{S}^{+}_{{\bf r}} \hat{S}^{+}_{{\bf r}+{\bf e}} \hat{S}^{-}_{{\bf r}^\prime}\hat{S}^{-}_{{\bf r}^\prime+{\bf e}}\rangle
\sim \langle \hat{f}^{+}_{{\bf r}} \hat{\bar{f}}^{+}_{{\bf r}+{\bf e}}\hat{f}^{-}_{{\bf r}^\prime} \hat{\bar{f}}^{-}_{{\bf r}^\prime+{\bf e}}\rangle,\nonumber
\end{equation} 
where the $\sim$ symbol implies that the two are equal modulo a phase. 
The fermion 4-point correlator is calculated using the Wick's decoupling and  approaches a constant  
 as  $|{\bf r}-{\bf r}^\prime|\rightarrow \infty$. Therefore, $C^{(4)} (\infty) = {\rm const}$ and we have an ordered state. This proves the existence of a spin-nematic-type long-range order, but does not prove (or rule out) the existence of a ``stronger'' magnetic order, which would require calculation of 2-point correlators.  

 
 {\it Symmetries and topology of the CS superconductor --} 
We now discuss  symmetries of the BdG Hamiltonian (\ref{Hamiltonian-2}). 
Importantly, it fulfills simultaneously {\em (i)} particle-hole (PH) symmetry, $\sigma_2 H^*_{BdG}\sigma_2= H_{BdG}$,
physical meaning of which is conservation of pseudo-spin linked with Fermi statistics; 
and {\em (ii)} time reversal (TR) symmetry 
$\tau_1 H_{BdG}^T\tau_1=- H_{BdG}$. Here, the Pauli matrix $\tau_1$  
acts in the space of $K$, $K^\prime$ blocks, which is also locked with the Nambu space. 
The combination of the effective PH and TR symmetries forms a chiral symmetry $\tau_1 \sigma_2 H_{BdG} \tau_1 \sigma_2= - H_{BdG}$, 
which defines the symmetry class $DIII$  for the Hamiltonian  Eq.~(\ref{Hamiltonian-2}) according to Altland-Zirnbauer classification [\onlinecite{Altland, Zirnbauer, Chiu}].



The presence of an effective chiral symmetry implies that one can choose a different basis,  where the BdG Hamiltonian decouples into two models  
having opposite chiralities.  Using a unitary transformation $\phi_{\bf k}=(\phi^L_{\bf k},\phi^R_{\bf k})= U \psi_{\bf k}$ we can reduce $H$ to a simpler form by introducing two copies of two component ``left'' (L) and ``right'' (R) fermions with  masses of opposite signs, $\pm \Delta_{3,\bf k}$:
\bea
\label{Hamiltonian-3}
U^+ H U=\left(
\begin{array}{cc}
H^L & 0\\
 0 & H^R
\end{array}
\right),\;\;
U=\frac{1}{\sqrt{2}}\left(
\begin{array}{ll}
\sigma_3& 1\\
-1&\sigma_3 
\end{array}
\right),
\ena
where $H^{R/L}=\pm \Delta_{3 \bf k}/2+ (\Delta_{0,\bf k}/k \mp v_F){\bf k} \cdot \boldsymbol{\sigma}$.
 Figure~\ref{gap-graphic} depicts the spectrum of the lower energy branch of R/L states.
Eq.~(\ref{Hamiltonian-3}) describes a superconducting  state,
which supports a pair of couter-propagating massless edge states, $R$ and $L$,  
 that are protected by the chiral (combination of TR and PH) symmetry. Note that  the  edge states
 are present not only in the theory with $\gamma=0$ (an isotropic spin model in the original language),
 but also at a finite $\gamma$. In fact, the existence of a CS superconducting state in the latter is more natural,
 in that it does not require spontaneous symmetry breaking.  Since the gauge symmetry  is broken from the outset by the non-zero $\gamma$, there aren't  Goldstone excitation and the corresponding {\em bulk} ground-state is truly gapped.


{\it Relation to SPT phases --} The existence of two counter-propagating Majorana edge modes in the Bogoliubov-de Gennes mean-field of the CS superconductor should
correspond to some low energy edge excitations in the original lattice spin model with a boundary. Specifically, the edge theory should be equivalent to the 1D critical quantum spin-1/2 Ising chain\cite{Volovik,Read,bern,fendley}  -- an example of 1+1D conformal field theory with central charge $c = (1/2, 1/2)$. The choice of spin-1/2 operators in the original Hamiltonian (\ref{H}) as $ \hat{\bf S}_{\bf r} = \frac{\hbar}{2}  \hat{\bm \sigma}_{\bf r}$, where $\hat{\sigma}^j_{\bf r},$ $j = x; y; z$, are the Pauli matrices, yields a topologically trivial, conventional XY magnet, $H = H_{XY} (\gamma)$, with the well-known  antiferromagnetic N\'{e}el  ground state. It does not exhibit any edge states. One can show that the reason is that the microscopic  edge Hamiltonian for would-be gapless modes of such a model, breaks the chiral symmetry  and thus the corresponding edge-states are actually gaped.   

Nevertheless this suggests that  CS $p$-wave superconductor  corresponds to the spin operators being {\em not} the Pauli matrices, but  ``twisted spin operators''  -- B-operators of Levin and Gu, mentioned in the introduction. Indeed, the Levin-Gu construction hinges on the existence of non-trivial element(s) of the group ${\mathcal H}^3 [\mathds{G}, U(1)]$ in  Wen's\cite{wen1,wen-sci,top-class} classification of SPT states. In the case of an Ising model on the triangular lattice, the group $\mathds{G}$ is the Ising $Z_2$ symmetry,  giving rise to two kind of phases, which can be associated with two types of spin operators - the Pauli matrices (yielding the trivial phase) and B-operators (yielding the non-trivial SPT phase). Similarly in our case of two copies of Levin-Gu-type models for each triangular sub-lattice, there is non-trivial element of ${\mathcal H}^3$ indicating the B-operators representation. The corresponding edge Hamiltonian complies with the bulk symmetries, keeping edge-modes gapless. 

Our construction of $p$-wave CS superconductor thus corresponds to such  new kind of a XY model that can be dubbed  twisted $XY$ or $TXY$ model. 
We notice that
this SPT state coexists with the long-ranged nematic (and possibly Neel) order, spontaneously breaking the $U(1)$ symmetry of the parent Hamiltonian (\ref{H}) at $\gamma=0$.   
One attribute the  $TXY$ model must have is that it should be possible to gap out its counter-propagating edge states on two sublattices and turn it into a conventional $XY$ magnet by breaking the  ``protecting'' symmetry, reproducing a topologically trivial state at the mean-field level.

In conclusion, we note that  gauging the Chern-Simons superconductor gives rise to the topologically ordered state (in the sense that it allows anyon excitations in the bulk). 
We also note that the corresponding state is different from the Moore-Read state (which corresponds to a gauged $p+ip$ superconductor, while our parent fermionic state preserves time-reversal symmetry). We defer a detailed discussion of this kind of topological order to future studies.

The authors are grateful to Andrew Allocca, Fiona Burnell, Michael Hermele, Olexei Motrunich, Andrew Potter, Zach Raines, and Kai Sun for useful discussions.
This work was supported by the PFC-JQI (T.S.), DOE-BES (DESC0001911) and Simons Foundation (V.G.), and DOE contract DE-FG02-08ER46482 (A.K., T.S.).
The work of T.S. and V.G. was performed in part at the Aspen Center for Physics, which is supported by National Science Foundation grant PHY-1066293.


\pagebreak
\widetext
\begin{center}
\textbf{\large Supplementary material for ``Topological spin ordering via Chern-Simons superconductivity''}
\end{center}
\setcounter{equation}{0}
\setcounter{figure}{0}
\setcounter{table}{0}
\setcounter{page}{1}
\makeatletter
\renewcommand{\theequation}{S\arabic{equation}}
\renewcommand{\thefigure}{S\arabic{figure}}
\renewcommand{\bibnumfmt}[1]{[S#1]}
\renewcommand{\citenumfont}[1]{S#1}

\author{Tigran Sedrakyan}
\affiliation{William I. Fine Theoretical Physics Institute,  University of Minnesota, Minneapolis, Minnesota 55455, USA}
\affiliation{Physics Frontier Center at the Joint Quantum Institute, University of Maryland, College Park, Maryland 20742, USA}

\author{Victor Galitski}

\affiliation{ Joint Quantum Institute and Condensed Matter Theory Center, Department of Physics,
University of Maryland, College Park, Maryland 20742-4111, USA}
\affiliation{School of Physics and Astronomy, Monash University, Melbourne, Victoria 3800, Australia}

\author{Alex Kamenev}
\affiliation{William I. Fine Theoretical Physics Institute,  University of Minnesota, Minneapolis, Minnesota 55455, USA}


In this Supplementary material, we present details on {\em (i)}
the derivation of gap equations~(\ref{gap-eq}) of the main text for  
superconducting order parameters $\Delta_{0 \bf k} $ and $ \Delta_{3 \bf k}$ and {\em (ii)} their asymptotic solution. 
We start with the saddle point equations 
Eq.~(\ref{gap}) that follow from minimization of action:
\bea
\label{S-gap}
\Delta^{\alpha\alpha'}_{\bf k}=
\sum_{\beta\beta' \bf k'}V^{\alpha\alpha',\beta\beta'}_{\bf k-k'}\langle \hat{\bar{f}}_{ -\bf k', \beta} \hat{f}_{ \bf k',\beta'}
\rangle.
\ena
The vacuum expectation value of Cooper pair annihilation operator  $\langle \hat{\bar{f}}_{ -\bf k', \beta} \hat{f}_{ \bf k',\beta'}\rangle $ 
can be represented as the derivative of the effective fermionized action $W(\{\Delta^{\alpha\alpha'}_{\bf k}\})=\sum_{{\bf k},a=\pm} E^{(a)}_{\bf k}$ 
with respect to Hubbard-Stratonovich fields $\Delta^{\beta\beta'}_{\bf k}$ as 
\bea
\label{cp}
\langle \hat{\bar{f}}_{ -\bf k', \beta} \hat{f}_{ \bf k',\beta'}\rangle 
=\frac{\delta W(\{\Delta^{\alpha\alpha'}_{\bf k}\})}{2 \delta\Delta^{\beta\beta'}_{\bf k}} .
\eea 
Substituting Eq.(\ref{cp}) into (\ref{S-gap}), one obtains 
\bea
\label{1}
\Delta^{\alpha\alpha'}_{\bf k}=\frac{1}{2}
\sum_{\beta\beta' \bf k'}V^{\alpha\alpha',\beta\beta'}_{\bf k-k'}
\frac{\delta W(\{\Delta^{\alpha\alpha'}_{\bf k}\})}{\delta\Delta^{\beta\beta'}_{\bf k}}.
\ena

Following the reasoning of the main text, we introduce the following convenient notations
\bea
\label{N}
 \Delta^{12}_{ \bf k}&=&- \Delta^{21}_{ \bf k}=\Delta^x_{\bf k}-i \Delta^y_{\bf k}=\Delta_{0 \bf k}k^-/k\nn\\
\Delta^{11}_{ \bf k}&=&- \Delta^{22}_{ \bf k}=\Delta_{3 \bf k},
\ena
where $k^\pm=k_x\pm i k_y$. Then, using the dispersion relation $E^{(a)}_{\bf k}=\sqrt{| a v_F {\bf k}+ \boldsymbol{\Delta}_{0 \bf k}|^2+|\Delta_{3 \bf k}|^2}$, it is straightforward to  
take the variation of the action $W(\{\Delta^{\alpha\alpha'}_{\bf k}\})$ with respect to the Hubbard-Stratonovich fields. Finally, we use explicit expression (\ref{HS}) of 
momentum dependent interaction vertices  $V^{\alpha\alpha',\beta\beta'}_{\bf k-k'}$, 
\bea
\label{V}
V^{12,11}_{\bf k-k'}&=&2 \pi e v_F A^-,\;\;
V^{21,11}_{\bf k-k'}=- 2 \pi e v_F  A^+\nn\\
V^{12,22}_{\bf k-k'}&=&2 \pi e v_F A^-,\;\;
V^{21,22}_{\bf k-k'}=- 2 \pi e v_F  A^+\\
V^{22,12}_{\bf k-k'}&=&-2 \pi e v_F A^+,\;\;
V^{11,12}_{\bf k-k'}=- 2 \pi e v_F  A^+\nn\\
V^{11,21}_{\bf k-k'}&=&2 \pi e v_F A^-,\;\;
V^{22,21}_{\bf k-k'}=2 \pi e v_F  A^-\nn,
\ena
to rewrite self-consistent gap equations (\ref{1}) in the following form:
\bea
\label{2}
\Delta_{0 \bf k}\frac{k^{-}}{k}&=&- \pi e v_F\sum_{{\bf k'},a=\pm } A^-_{{\bf k}-{\bf k'}}\frac{\Delta_{3 \bf k'}}{ E^{(a)}_{\bf k'}},\nn\\
\Delta_{3 \bf k}&=&\frac{\pi e v_F}{2}\sum_{{\bf k'},a=\pm }\frac{A^-_{{\bf k}-{\bf k'}}k^{\prime +}+A^+_{{\bf k}-{\bf k'}}k^{\prime-}}{k'} 
\frac{ \Delta_{0 k'}+a v_F k'}{ E^{(a)}_{\bf k'}}.
\ena
Here, ${\bf A_q}={\bf q}/|{\bf q}|^2 $ is the Fourier image of the  vector potential of the vortex gauge field 
${\bf A}({\bf r})$ , and $A^{\pm}_{\bf q}=A^{x}_{\bf q}\pm i A^{y}_{\bf q} $. 

Eqs.~(\ref{2}) can be simplified further, namely integration over the relative angle $\phi$ between ${\bf k}$
and ${\bf k'}$ vectors can be performed analytically. This task can be accomplished using the following identities:
\bea
\label{3}
\int_0^{2\pi}\frac{d\phi}{2\pi} \frac{(\bf k -k'){\bf k'}}{(\bf k- k')^2}= \int_0^{2\pi} \frac{d\phi}{2\pi} \frac{k- k' \cos[\phi]}{k^2+k^{' 2}-2 k k' \cos[\phi]}=\frac{1}{k} \theta[k-k']\nn\\
\int_0^{2\pi} \frac{d\phi}{2\pi} \frac{k- k' e^{i \phi}}{k^2+k^{' 2}-2 k k' \cos[\phi]}=\frac{1}{k} \theta[k-k'].
\ena
After performing the angular integration in Eqs.~(\ref{2}),  one obtains    
\bea
\label{S-gap-eq}
\Delta_{0 \bf k}&=&\frac{e v_F}{2}\sum_{a=\pm }\int_0^{k} dk' \frac{k' \Delta_{3 k'}}{k E^{(a)}_{\bf k'}},\nn\\
\Delta_{3 \bf k}&=&\frac{e v_F}{2}\sum_{a=\pm }\int_{k}^{\Lambda} dk' \frac{ \Delta_{0 k'}+a v_F k'}{ E^{(a)}_{\bf k'}}.
\ena 
In this way, one reproduces gap equations of the main text. Below we will discuss the asymptotic solution of these equations. 

\begin{figure}[t]
\includegraphics[width=100mm,angle=0,clip]{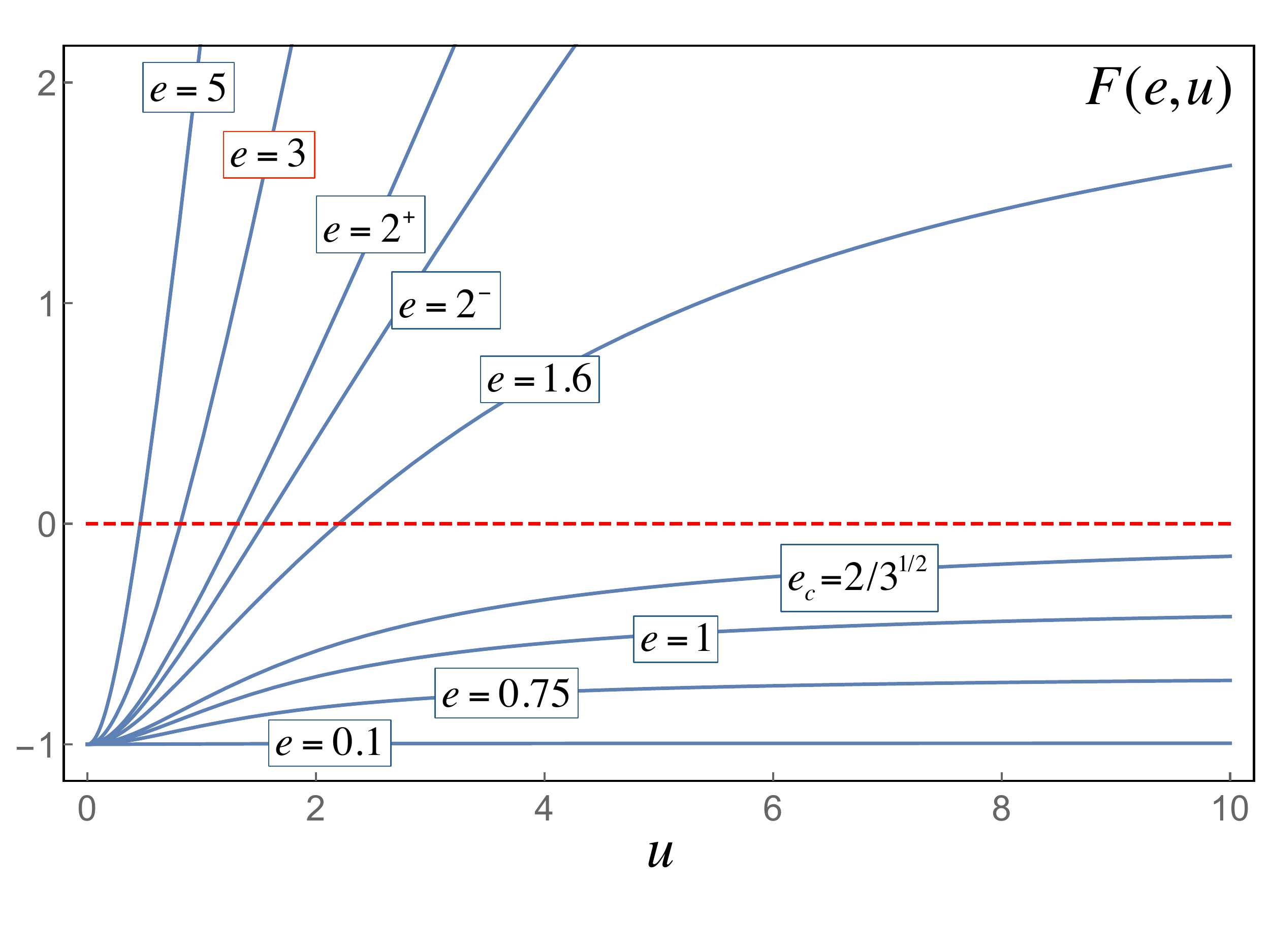}
\caption{(Color online) The function $F(e,u)$ is plotted vs $u$ for various values of $e$.  At 
 $e>e_c=2/\sqrt{3}$ the function $F(e,u)$ crosses the dashed line 
giving raise to a solution to $F(e,u)=0$ equation. Exactly at the critical point $e=e_c$ the  function $F(e_c,u)$ asymptotically approaches the dashed line from below.  
} 
\label{gap2}
\end{figure}

In the limit $k\rightarrow 0$ it is seen upon inspecting the first Eq.~(\ref{S-gap-eq}) that $\Delta_{0 k\rightarrow 0}\simeq ev_Fk/2+O(k)^3$ and 
$ \Delta_{3 k\rightarrow 0}\equiv\Delta_{3 } = const\cdot (v_F\Lambda)+O(k)^2$ is indeed a solution of it. The self-consistency requires however that this asymptotic solution should 
also satisfy the second  Eq.~(\ref{S-gap-eq}). Substituting $\Delta_{0 k\rightarrow 0}$ and $ \Delta_{3 }$ into the latter, and using the form of the spectrum $E^{(a)}_{\bf k}$, one obtains 
\bea
\label{second-gap}
\Delta_{3}=\frac{ev_F}{2}\int_{0}^{\Lambda} dk'\left[\frac{(\frac{e}{2}+1)v_Fk'}{\sqrt{\left(\frac{e}{2}+1\right)^2(v_Fk')^2+\Delta_3^2}}+\frac{(\frac{e}{2}-1)v_Fk'}{\sqrt{\left(\frac{e}{2}-1\right)^2(v_Fk')^2+\Delta_3^2}}\right].
\eea
Integration over $k'$ in Eq.~(\ref{second-gap}) can be readily performed. Upon introducing a new dimensionless variable $u=(\Lambda v_F)/\Delta_3$, Eq.~(\ref{second-gap}) assumes the simple algebraic form:
\bea
\label{form}
F(e,u)=0,
\eea
where
\bea
\label{F}
 F(e,u) = \frac{e}{2}\left[\frac{1}{\frac{e}{2}+1}\left(\sqrt{\left(\frac{e}{2}+1\right)^2 u^2+1}-1\right)+ \frac{1}{\frac{e}{2}-1}\left(\sqrt{\left(\frac{e}{2}-1\right)^2 u^2+1}-1\right)\right]-1.
\eea
The function $F(e,u)$ is plotted vs $u$ for various values of $e$ in Fig.~\ref{gap2}. We see that the solution to Eq.~(\ref{form}) (and thus to self-consistent gap equations Eq.~(\ref{S-gap-eq})) exists only for $e>e_c=2/\sqrt{3}$. Such a phase transition at $e_c$ can be seen from large and small $u$ asymptotes of function $F$. These are given by
\bea
\label{asymptotes}
&&F(e,u\gg1)=\frac{eu}{2}\left[1+ \text{sgn}(e-2)\right]+\frac{3e^2-4}{4-e^2}+O(1/u), \nn\\
&&F(e,u\ll1)=-1+\frac{e^2u^2}{4}+O(u)^4.
\eea
We see that at $e<e_c=2/\sqrt{3}$, the monotonically increasing function $F(e,u)$ at $u\rightarrow\infty$ asymptotically approaches a negative constant value, $F(e,\infty)=(3e^2-4)/(4-e^2)$, implying that there is no solution to Eq.~(\ref{form}) in this region.    
The large-$u$ behavior of $F(e,u)$ is the same at $2> e >e_c$, but in this region the constant $F(e,\infty)$ is positive,
and thus $F(e,u)$ passes through zero (notice that $F(e,0)=-1$), giving raise to a solution to Eq.~(\ref{form}) at some finite $u$.  For $e>2$, the linear 
large-$u$ asymptote sets in and the condition Eq.~(\ref{form}) is being satisfied at even smaller $u$.
At $e=3$, the equation $F(3,u)=0$ has a solution $u=u_0=0.814$. This means that for physical value $e=3$, the order parameter acquires the asymptotic form $\Delta_{3k\rightarrow 0}\simeq1.23 \Lambda v_F+O(k)^2$.


\end{document}